\documentclass[11pt]{article}
\usepackage{multicol}
\usepackage{amsmath,amsfonts,amssymb,amsthm,array}
\usepackage[margin=1in]{geometry}
\usepackage[colorlinks]{hyperref}
\usepackage{bbold}
\usepackage{titlesec}
\usepackage[toc,page]{appendix}

\usepackage{mathtools}
\DeclarePairedDelimiter\bra{\langle}{\rvert}
\DeclarePairedDelimiter\ket{\lvert}{\rangle}
\DeclarePairedDelimiterX\braket[2]{\langle}{\rangle}{#1 \delimsize\vert #2}

\usepackage{nicefrac}
\usepackage{amsfonts}
\usepackage{footmisc}
\usepackage{amsmath}
\usepackage[]{titletoc}
\usepackage{titlesec}
\usepackage[T1]{fontenc}
\usepackage[utf8]{inputenc}
\usepackage{amssymb}
\usepackage{fancyhdr}
\usepackage{wrapfig}
\usepackage{caption} 
\usepackage{amssymb,amsmath,xcolor,graphicx,xspace,colortbl, rotating} 
\usepackage{hyperref}  
\usepackage{transparent}
\usepackage[utf8]{inputenc}
\usepackage[T1]{fontenc}
\usepackage{fancyhdr}
\usepackage{eso-pic}
\usepackage{graphicx}
\usepackage{mathtools, bm}
\usepackage{amssymb, bm}
\usepackage{subcaption}
\usepackage{setspace}
\usepackage{multicol}
\usepackage{fancybox}
\usepackage{hyperref}
\usepackage{emptypage}
\usepackage{epstopdf}
\usepackage{float}
\usepackage[]{placeins}
\providecommand{\U}[1]{\protect\rule{.1in}{.1in}}

\begin{document}


\title{On the Negativity of Wigner Function as a measure of entanglement under quantum  polarization converter devices}

\author{ Mustapha Ziane\footnote{mustapha-1231@hotmail.com} \hspace{0.1cm} and\hspace{0.1cm} Morad El Baz\footnote{modar.elbaz@fsr.ac.ma}\\Université Mohammed V, Faculté des Sciences\\
Equipe Sciences de la Matière et du Rayonnement\\
Av. Ibn Battouta, B.P. 1014, Agdal, Rabat, Morocco}

\date{\today}
\maketitle

\begin{abstract}
We study the behaviour of the Negativity of Wigner Function (NWF) as a measure of entanglement in non-Gaussian states under quantum polarisation converter devices. We analyze comparatively this quantity with other measures of entanglement in a system prepared in a superposition of two-mode coherent states. We show that the (WF) can be identified as a quantifier of non-Gaussian entanglement.
\end{abstract}

\section*{Keywords}  Wigner function, Q-function, Negativity, Non-Gaussian state, Non-gaussianity, Coherent states, Quantum polarization. 

\section*{Introduction}
\label{sec1}
Quantum entanglement, is recognised as one of  quantum resources for several applications in quantum information theory including quantum computing, quantum communications and quantum cryptography. However, quantifying quantum correlations particularly, entanglement is one of the most relevant challenges in quantum information theory.\\

On the other hand polarization play a central role in a large number of optical phenomena. It appear in several paradigmatic application including remote sensing and light scattering \cite[3 abdo]. In recent years, a considerable attention has been paid to the polarization  of quantum light due to its application in quantum information protocols since the light can be extensively used in the quantum information-coding. 

There is a wide consensus that coherent states are the \textit{most classical} quantum states. It has been shown that these states minimize the Heisenberg uncertainty relation for position and momentums operators. In addition the dynamic of their expectation values has the same form as their classical counterpart. These properties, make coherent states least quantum states. For this reason, they are called quasi-classical states. These states describe coherent optical light and  can be generated in the  laboratory. In this way, the quantum polarization formalism is extended from  that  of classical light polarization through replacing the Stokes parameters  by the  associated stocks operators.

In addition, the degree of polarization of classical light does not depend on its intensity. However, the degree of polarization of coherent states decreases with decreasing values of optical power. Furthermore, the quantum Stokes parameters $\hat{S}_{1}, \hat{S}_{2}$ and $\hat{S}_{3}$ do not commute with each other. Then, it is not possible to know the values of any two of them simultaneously without uncertainties. This fact has been used in \cite{vidiella2006continuous} in order to create a continuous variable quantum key distribution system.
Basically, polarization of coherent states has extensively been used in quantum information processing, precisely in quantum key distribution protocols with continuous variables. \cite{vidiella2006continuous,barbosa2003fast,yin2007decoy,
kye2005quantum}.\\
Recently, one of the most important applications of the Wigner function in quantum information theory is the classification of classical and non-classical states based on its non-Gaussianity and its negativity \cite{kenfack2004negativity,genoni2007measure}. In fact, a Gaussian state is associated with a
Gaussian Wigner function in phase space of one mode or multimode pure state in continuous variables systems \cite{wenger2004non}.  On the other hand, a non-Gaussian state with negative Wigner function reveals that the system possesses non-classical correlations \cite{braunstein1998teleportation}. Recent works showed that the negativity of the Wigner function can detect the entanglement and it is not sensitive to all kinds of quantum correlations and can be a best quantifier of genuine entanglement in tripartite systems  \cite{ziane2018direct,ziane2018NWF}.\\
In this work, we study the strength  of the NWF as a quantifier of entanglement under a quantum polarization converter devices. In this direction we describe in section \eqref{sec2} the Stokes  parameters in order to analyse the quantum polarization of the superposition of two bi-mode coherent states; As well in section \eqref{wignerentanglement}, we use the entanglement of formation and the NWF to analyse the entanglement behavior before and after  the polarization converter devices; Finally, in section \eqref{discussion} we discuss our results and provide conclusions.%
\section{Review of quantum polarization }\label{sec2}
In classical optics the polarization of light beams is determined by computing the four Stokes parameters \cite{stokes1852gg}. Similarly, In quantum optics the degree of polarization of states of light is quantified by the calculation of the mean values of the associated Stocks operators\cite{klimov2005distance,luis2002degree}.\\For a monochromatic plane wave propagating in the $z$-direction, whose electric field lines in the $xy$ plane. In terms of the annihilation and creation operators of horizontally and vertically polarized modes noted by $\hat{a}_{H}$ and $\hat{a}_{V}$, respectively, the Stokes operators can be expressed as\cite{klimov2005distance,luis2002degree}
\begin{equation}
\begin{matrix}
\hat{S}_0 &=&\hat{a}_H^+\hat{a}_H+\hat{a}_V^+\hat{a}_V,\\
\hat{S}_1 &=&\hat{a}_H^+\hat{a}_H-\hat{a}_V^+\hat{a}_V,\\
\hat{S}_2 &=&\hat{a}_H^+\hat{a}_V+\hat{a}_V^+\hat{a}_H,\\
\hat{S}_3 &=&i\left(\hat{a}_V^+\hat{a}_H+\hat{a}_H^+\hat{a}_V\right)
\end{matrix}
\end{equation}and the Stokes parameters are given by calculation of the corresponding average values $<\hat{S}_k>$. Using the bosonic commutation relations
\begin{equation}
\left[\hat{a}_i,\hat{a}_j^+\right]=\hat{\textbf{1}}\delta_{ij} \quad;\quad \lbrace i,j\rbrace\in\lbrace H,V\rbrace.
\end{equation} It has been shown that, the operators $\hat{S}_1$, $\hat{S}_2$ and $\hat{S}_3$ all commute with $\hat{S}_0$ and satisfy the corresponding commutation relations,
\begin{equation}\label{commu}
\left[\hat{S}_k,\hat{S}_l\right]=2i\hat{S}_m \quad;\quad\lbrace k,l,m\rbrace \in\lbrace1,2,3\rbrace.
\end{equation} 
The Stokes operators $\hat{S}_1$, $\hat{S}_2$ and $\hat{S}_3$ thus from $SU\left(2\right)$ algebra and  generate all transformations from this group;
\begin{itemize}
\item $\hat{S}_2$ is the infinitesimal generator of geometric rotations around the direction of propagation,
\item $\hat{S}_3$ is the differential phase shifts between the two modes.
\end{itemize}

For a quasi-classical two-mode coherent state $\ket{\alpha,\beta}$ defined as 
\begin{equation}\label{pcs}
\ket{\alpha,\beta}=\mathrm{e}^{-\dfrac{\lvert \alpha\rvert +\lvert\beta\rvert}{2}}\sum_{n,m}\dfrac{\left(\alpha\right)^n}{\sqrt{n!}}\dfrac{\left(\beta\right)^m}{\sqrt{m!}}\ket{n,m},
\end{equation}
the mean values $<\hat{S}_i>$ of the three Stokes operators and the variances $V_i$  are expressed as a function of the field amplitudes:
\begin{equation}
\begin{matrix}
<\hat{S}_1>=&\lvert\alpha\rvert^2- \lvert\beta\rvert^2,&<\hat{S}_1^2>=&\left(\left(\lvert\alpha\rvert\right)^2-\left(\lvert\beta\rvert\right)^2\right)^2+\left(\lvert\alpha\rvert\right)^2+\left(\lvert\beta\rvert\right)^2,\\

<\hat{S}_2>=&\alpha^*\beta-\alpha\beta^*,\quad& <\hat{S}_2^2>=&\left(\alpha^*\beta\right)^2+\left(\alpha\beta^*\right)^2+\lvert\alpha\rvert^2+\lvert\beta\rvert^2 +2\lvert\alpha\rvert^2\lvert\beta\rvert^2,\\

<\hat{S}_3>=&i\left(\alpha\beta^*-\alpha^*\beta\right)  ,\quad& <\hat{S}_3^2>=&-\left(\alpha^*\beta\right)^2-\left(\alpha\beta^*\right)^2+\lvert\alpha\rvert^2+\lvert\beta\rvert^2 +2\lvert\alpha\rvert^2\lvert\beta\rvert^2,\\ 
\end{matrix}
\end{equation} and 
\begin{equation*}
V_1=V_2=V_3=\lvert\alpha\rvert^2+ \lvert\beta\rvert^2.
\end{equation*}
The average values of the quantum Stokes parameters of coherent state are equal to of classical light Stokes parameters values and the variance of the three parameters increase while the optical power increases.
\section{Quantum degree of polarization of superposition of two mode coherent states} \label{qpola}
\begin{figure}
\centering
\begin{minipage}{.5\textwidth}
  \includegraphics[width=0.9\linewidth]{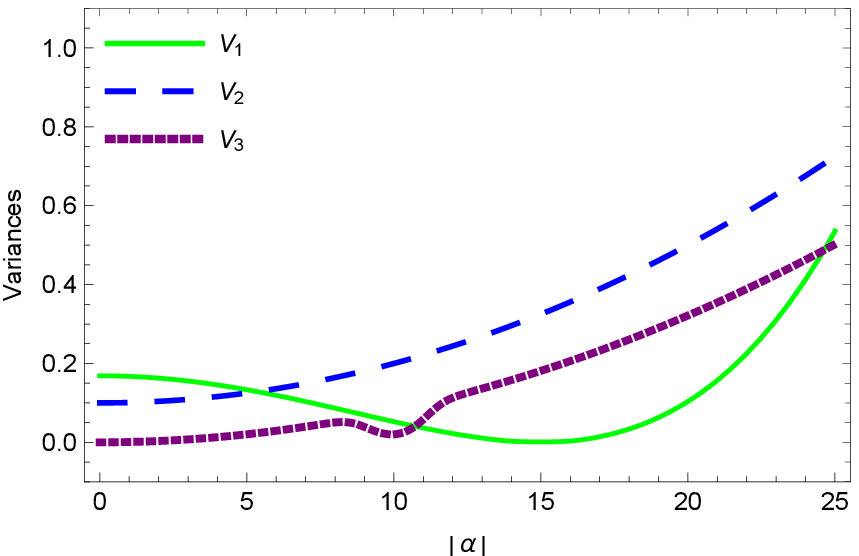}
  \captionof{figure}{Variances $V_1$, $V_2$, and $V_3$ of \\
  $\hat{S}_1$, $\hat{S}_2$ and $\hat{S}_3$ (respectivly) versus\\ $\lvert\alpha\rvert^2$ for $\ket*{\psi_1} $ having $\lvert\beta\rvert^2=4$.}
\label{var}
\end{minipage}%
\begin{minipage}{.5\textwidth}
  \centering
  \includegraphics[width=0.9\linewidth]{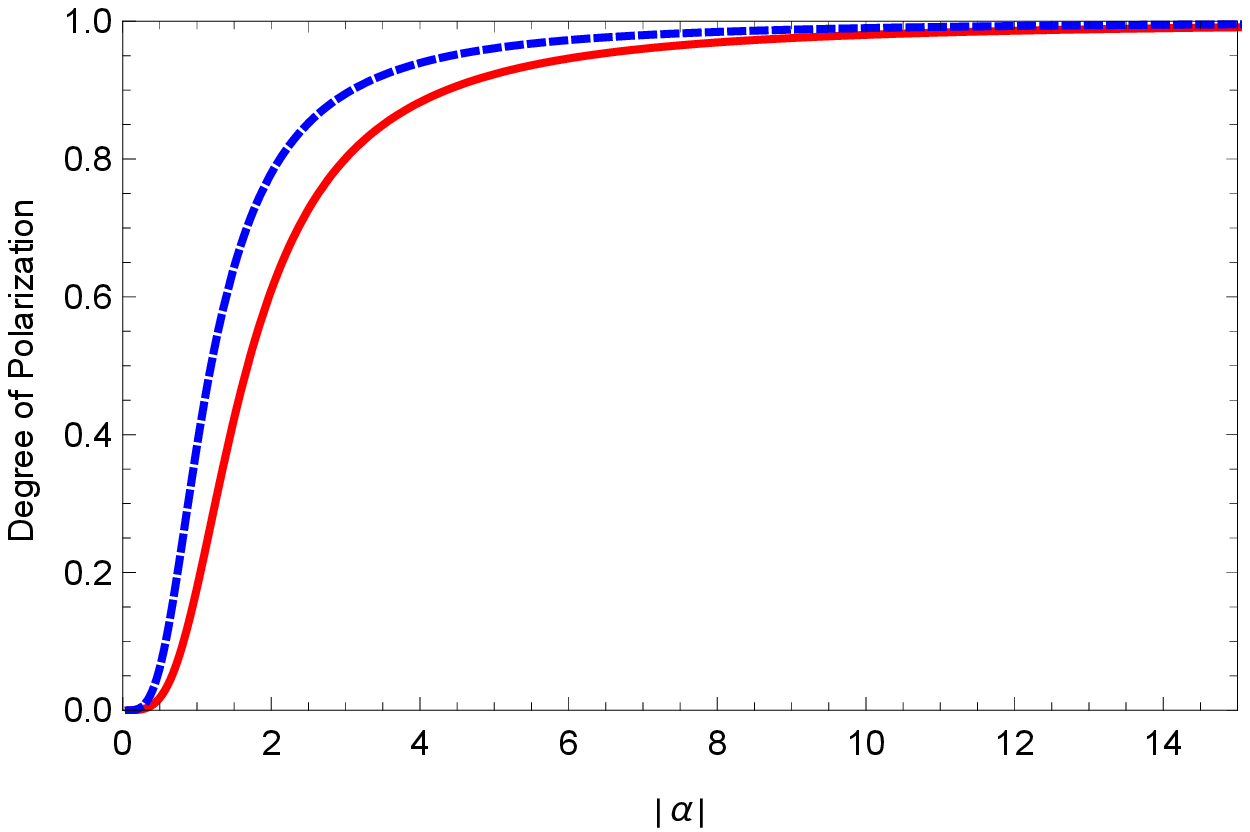}
  \captionof{figure}{Quantum degree of polarization for the states $\ket*{0,\pm\alpha}$  (Solidligne) and  $\ket*{\pm\alpha,\mp\alpha}$ (Dashedligne).}
\label{pola1}
\end{minipage}
\end{figure}

Classically, the light is considered unpolarized if it's Stokes parameters vanish. In quantum mechanics, this is condition necessary but not sufficient. A quantum light beam can be considered unpolarized if its observable do not change after a geometric rotation and/or a phase shift between the components. These condition are mathematically described by \cite{agarwal1996invariances}:
\begin{equation}
\left[\hat{\rho},\hat{S}_1\right]=\left[\hat{\rho},\hat{S}_3\right]=0,
\end{equation} 
where $\hat{\rho}$ is the density matrix of the quantum state.\\
By analogy with the classical definition, many measures of the quantum polarization degree have been proposed \cite{luis2002degree,klimov2005distance}. Here we consider a measure based on Q-function \cite{luis2002degree}:
\begin{equation}\label{degree}
P=\dfrac{D}{1+D}
\end{equation}
with 
\begin{equation*}
D=4\pi\int_0^{2\pi}d\Omega\int_0^{\pi}\left[Q\left(\theta,\phi\right)-\dfrac{1}{4\pi}\right]^2\sin\left(\theta\right)d\theta d\phi 
\end{equation*}
where $d\Omega=\sin\left(\theta\right)d\theta d\phi$ is the differential of solid angle and $Q\left(\theta,\phi\right)$ is the Q-function of the light. For the two-mode coherent state $\ket{\alpha e^{i\phi_\alpha},\beta e^{i\phi_\beta}}$ the Q-function reads \cite{viana2005mixture}:
\begin{equation}\label{qfunction}
Q\left(\theta,\phi\right)=\dfrac{e^{-\left(\lvert\alpha\rvert^2+\lvert\beta\rvert^2\right)}}{4\pi}\left(1+z\right)e^2
\end{equation}
where \begin{align*}
 z & = \left[\lvert\alpha\rvert\cos\left(\dfrac{\theta}{2}\right)\cos\left(\phi_\alpha+\phi\right)+\beta\sin\left(\dfrac{\theta}{2}\right)\cos\left(\phi_\beta\right)\right]^2\\
   &+\left[\lvert\alpha\rvert\cos\left(\dfrac{\theta}{2}\right)\sin\left(\phi_\alpha+\phi\right)+\beta\sin\left(\dfrac{\theta}{2}\right)\sin\left(\phi_\beta\right)\right]^2.
\end{align*}Using \eqref{degree} and \eqref{qfunction}, the quantum polarization degree of the quantum state $\ket{\alpha,0}$ is\begin{equation}\label{pdegree}
P=1-\dfrac{4\lvert\alpha\rvert^2}{1+2\lvert\alpha\rvert}.
\end{equation}
When $\rvert\alpha\lvert^2\gg1$, equation \eqref{pdegree} can be approximated by 
\begin{equation}
P\simeq 1-\dfrac{2}{\rvert\alpha\lvert^2}
\end{equation}Showing that the quantum degree of  polarization of two-mode coherent state increase with increasing of the light power. In Figure \eqref{pola1} we show the quantum degree of polarization of the states $\ket*{0,\pm\alpha},$  $\ket*{\pm\alpha,0},$ $\ket*{\pm\alpha,\mp\alpha}$ and $\ket*{\pm\alpha,\pm\alpha}$ that are equivalent to vertical polarization $\ket*{V}$, horizontal polarization $\ket*{H}$, anti-diagonal polarization $\ket*{-\dfrac{\pi}{4}}$ and diagonal polarization  $\ket*{\dfrac{\pi}{4}}$, respectively.\\
In Figure \eqref{pola1} the diagonal states have a larger quantum degree of polarization, compared with horizontal and vertical states, because they have a larger mean photon number (in total).\\
In order to discuss the quantum degree of polarization, we consider the quantum Stokes parameters of quantum state composed by the superposition of bimodal coherent states defined by
\begin{equation}\label{super}
\ket*{\psi_\pm}=N\left(\ket*{\alpha,\beta}\pm\ket*{\gamma,\lambda}\right)
\end{equation}
where $\rvert N\lvert^2=\left\{2
+\left(\zeta+\zeta^*\right)\emph{Exp}\left[-\left(\lvert\alpha\rvert^2+\lvert\beta\rvert^2+\lvert\gamma\rvert^2+\lvert\lambda\rvert^2\right)\slash 2 \right]\right\}^{-1}$ and $\zeta=\emph{Exp}\left(\alpha^*\gamma+\beta^*\lambda\right)$. The average of the quantum Stokes parameters and of their squared values and the Q-function of the state \eqref{super} are given in Appendix \eqref{average}.\\
For the raison of simplification, we choose to consider the following particular cases of the state \eqref{super},
\begin{subequations}\label{state}
\begin{align}
\ket*{\psi_1}=&N_1\left(\ket*{\alpha,\beta}+\ket*{\beta,\alpha}\right)\label{s1}\\
\ket*{\psi_2}=&N_2\left(\ket*{-\alpha,-\alpha}+\ket*{\alpha,\alpha}\right)\label{s2}\\
\ket*{\psi_3}=&N_3\left(\ket*{\alpha,0}+\ket*{0,\alpha}\right)\label{s3}
\end{align}
\end{subequations}with $N_1=\left\{2\left[1+\textnormal{Exp}\left(2\alpha\beta-\lvert\alpha\rvert^2-\lvert\beta\rvert^2\right)\right]\right\}^{-1\slash2}$, $N_2=\left\{2\left[1+\textnormal{Exp}\left(-4\lvert\alpha\rvert^2\right)\right]\right\}^{-1\slash2}$ and  $N_3=\left\{2\left[1+\textnormal{Exp}\left(-\lvert\alpha\rvert^2\right)\right]\right\}^{-1\slash2}$. The averages and the covariances of the quantum Stokes parameters of the states $\ket*{\psi_1}$, $\ket*{\psi_2}$ and $\ket*{\psi_3}$ are calculated from their general expressions of superposed bi-mode coherent state and given in  Appendix \eqref{average}. we found that $<\hat{S}_1>$ and $<\hat{S}_3>$ vanish for the states \eqref{s1}, \eqref{s2} and \eqref{s3}. That is consequence of the fact that the average of optical powers in horizontal and vertical polarizations are equal for these cases.
In the separable bi-mode state given in \eqref{pcs}, the Stockes parameters and the variances expressed propationally to optical power contrary to the superposed entangled state defined in \eqref{super}. The entangled state gives an interesting behavior of the variances of the Stokes parameters. The variances can increase and decrease when the total mean photon number increases, as are shown in Figure \eqref{var}.

\section{Wigner function and entanglement under polarization converter devices }\label{wignerentanglement}
\begin{figure}
\centering
\begin{minipage}{.5\textwidth}
  \includegraphics[width=0.9\linewidth]{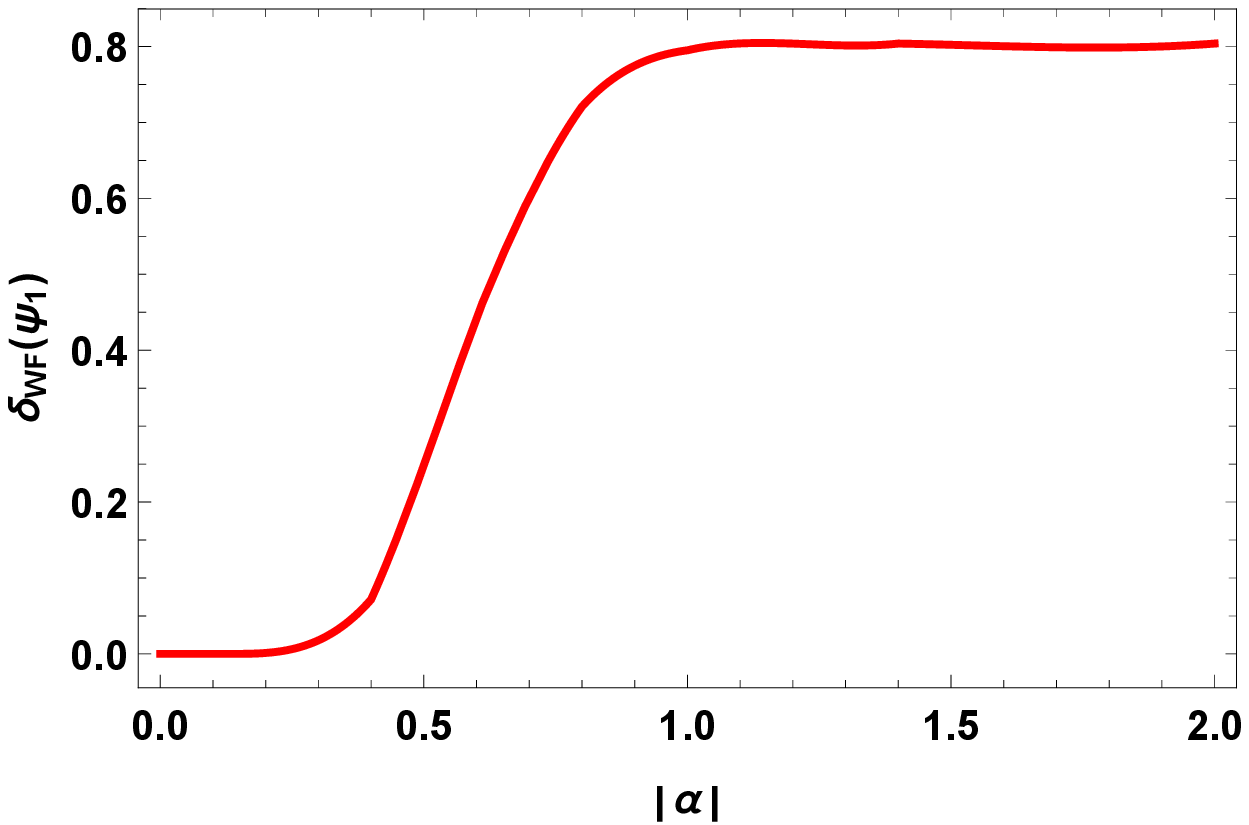}
  \captionof{figure}{The NWF  of the state $\ket*{\psi_1}$ \\versus $\lvert\alpha\rvert$ for  $\lvert\beta\rvert=2$.}
\label{negplott}
\end{minipage}%
\begin{minipage}{.5\textwidth}
  \centering
  \includegraphics[width=1\linewidth]{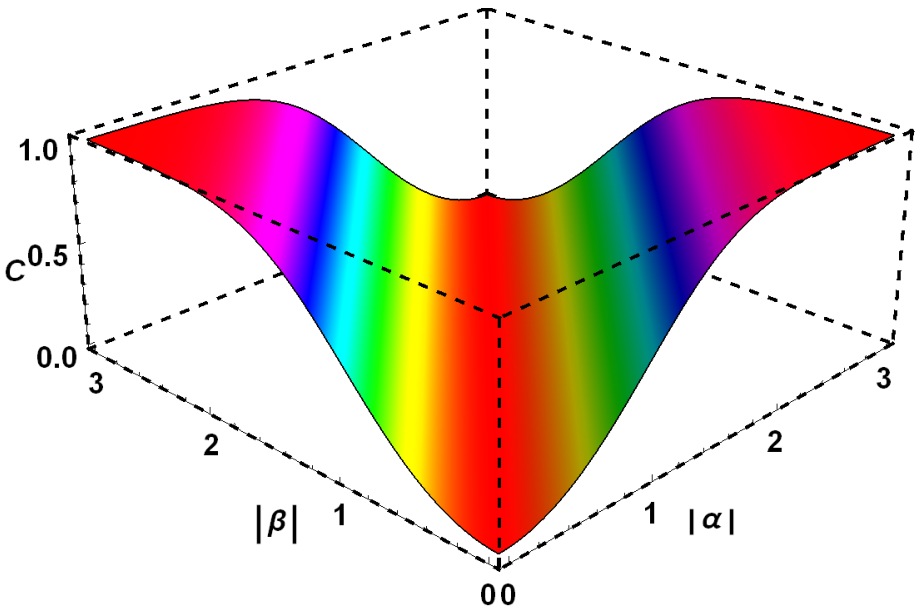}
  \captionof{figure}{The concurrence of the state $\ket*{\psi_1}$ versus $\lvert\alpha\rvert$ and  $\lvert\beta\rvert$.}
  \label{concplot}
\end{minipage}
\end{figure}

Wigner function is an important tool in physics, especially in quantum physics to detect the non-classical behavior of light  by studying its negativity. Recently, in quantum information theory the NWF is used as a measure of entanglement\cite{ziane2018direct}. In this section we study the NWF and the entanglement behavior of the superposition of  two-mode coherent states before and after pass by a polarization converter devices, in order to testing the strength of the NWF as a measure of entanglement.\\
For a single  quantum system described by a density matrix $\hat{\rho}$, the associated  Wigner function is defined by 
\begin{equation}\label{w}
\mathcal{W} \left( q,p\right) =\dfrac{1}{2\pi}\int \exp{(\dfrac{-ipy}{\hbar})}\bra*{q+\dfrac{y}{2}}\hat{\rho}\ket*{q-\dfrac{y}{2}}dy,
\end{equation}
where $\ket{q\pm\dfrac{y}{2}}$ are the eigenkets of the position operator. If the state in question is a pure state $\hat{\rho}=\ket{\psi}\bra{\psi}$ then
\begin{equation}
\mathcal{W} \left( q,p\right)= \dfrac{1}{2\pi\hbar} \int\psi^*\left(q-\dfrac{y}{2}\right)\psi\left(q+\dfrac{y}{2}\right)\exp{(\dfrac{-ipy}{\hbar})}dy.
\end{equation}
Hence, the doubled volume of the integrated negative part of the Wigner function may be written as \cite{kenfack2004negativity}
\begin{equation}
\delta\left(\rho\right)= \iint\left|\mathcal{W}\left(q,p\right) \right|\,dqdp-1.
\end{equation}
\begin{figure}

\begin{subfigure}{.5\textwidth}
  \includegraphics[width=.9\linewidth]{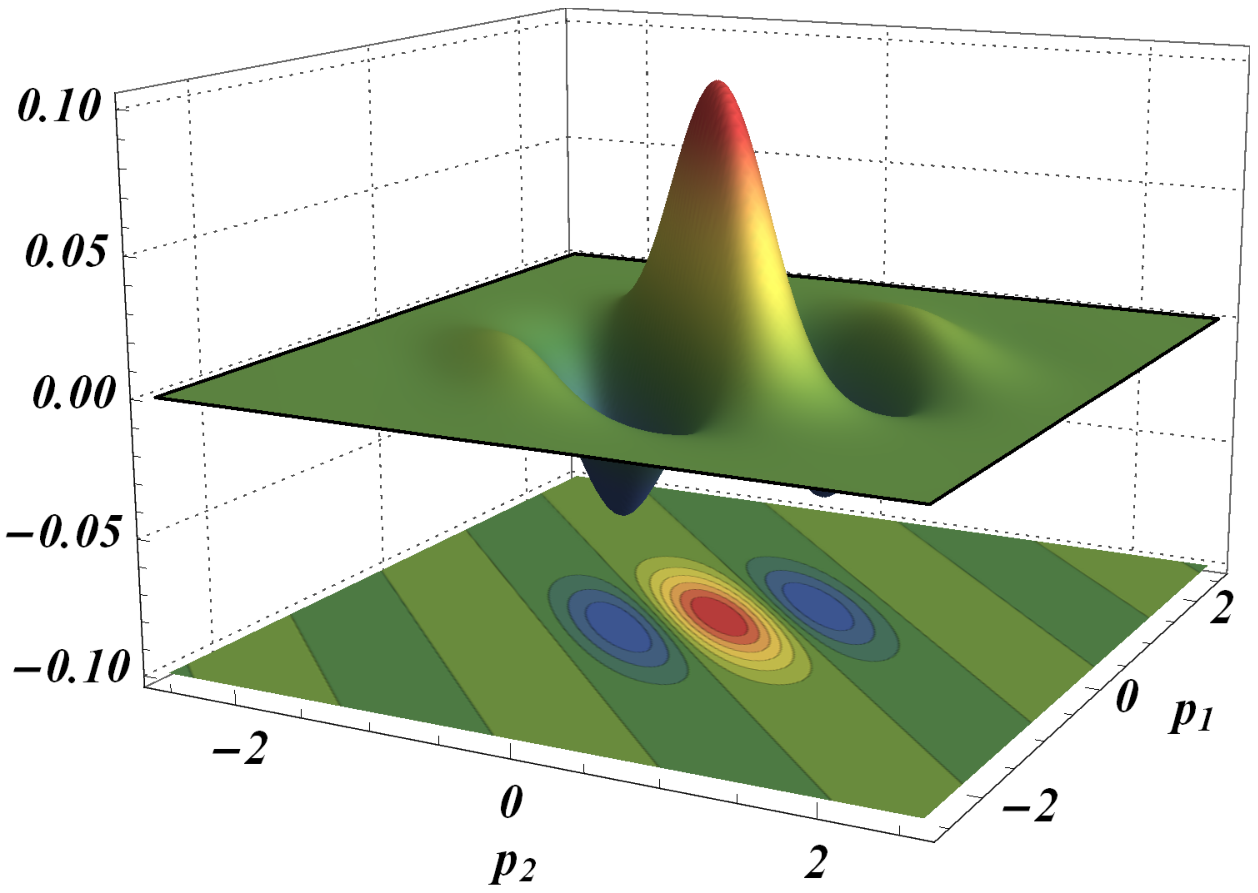}
  \caption{Wigner function of $\psi_1$ for $\left|\alpha\right|=\left|\beta\right|=1$.}
\end{subfigure}
\begin{subfigure}{.5\textwidth}
  \includegraphics[width=.9\linewidth]{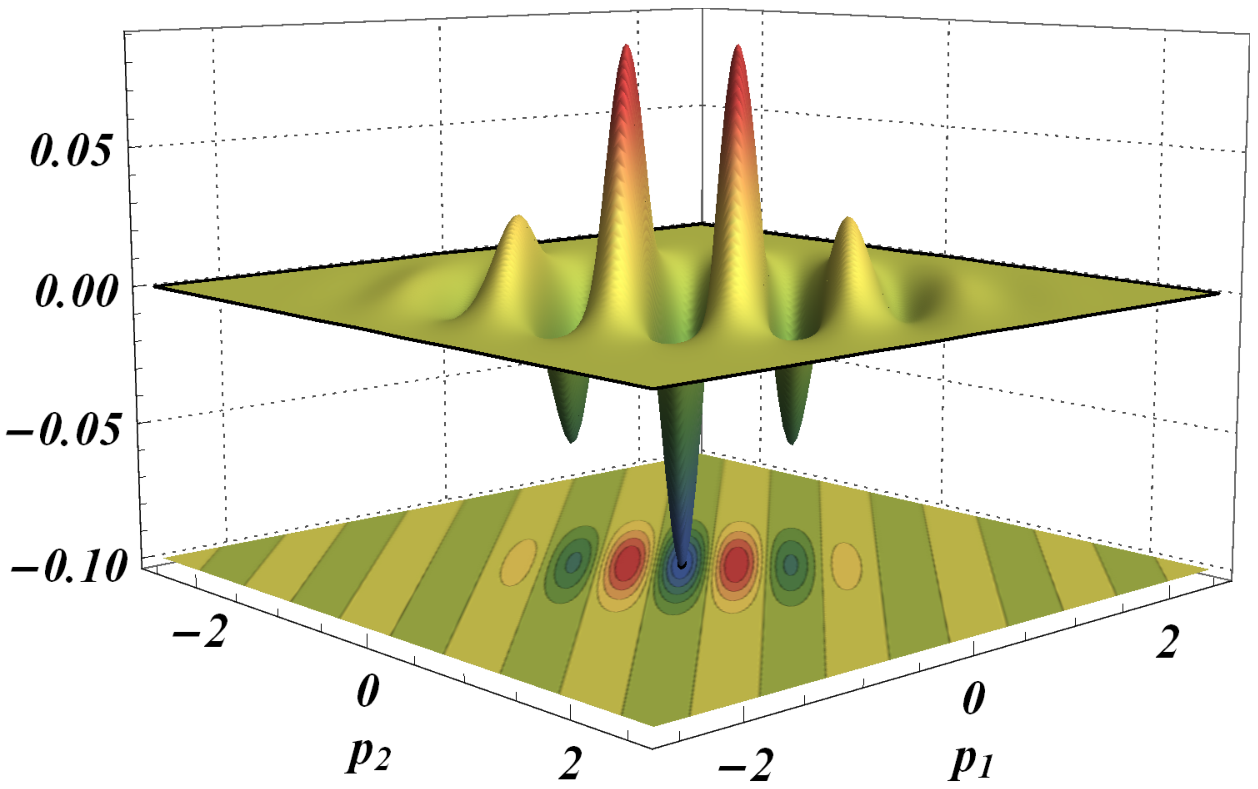}
  \caption{Wigner function of $\psi_+$ for $\left|\alpha\right|=\left|\beta\right|=2$.}
\end{subfigure}
\caption { Wigner function  of superposed coherent state (\eqref{s1}). } 
\label{wignerplot1} 
\end{figure}
It is clear from the plot in Figure \eqref{wignerplot1} that the Wigner function of the bi-mode superposed state \eqref{s1}  is not positive on the all phase space. The volume of the negative part of the Wigner function is plotted in Figure \eqref{negplott} versus $\lvert\alpha\lvert^2$ for $\lvert\beta\lvert=2.$
The quantum entanglement of the same state (\eqref{super}) can be measured by the concurrence \cite{rungta2001universal,kuang2003generation}
\begin{equation}
C=\dfrac{\sqrt{\left(1-\lvert\braket{\alpha}{\gamma}\rvert^2\right)\left(1-\lvert\braket{\lambda}{\beta}\rvert^2\right)}}{1+Re\left(\braket{\alpha}{\gamma}\braket{\beta}{\lambda}\right)}.
\end{equation}
The concurrence $C_{\psi_1}\left(\alpha\right)$ of  the particular state $\ket*{\psi_1}$ defined in \eqref{s1} is showen in Figure \eqref{concplot} .

\begin{figure}[H]
\begin{subfigure}{.5\textwidth}
  \includegraphics[width=0.94\linewidth]{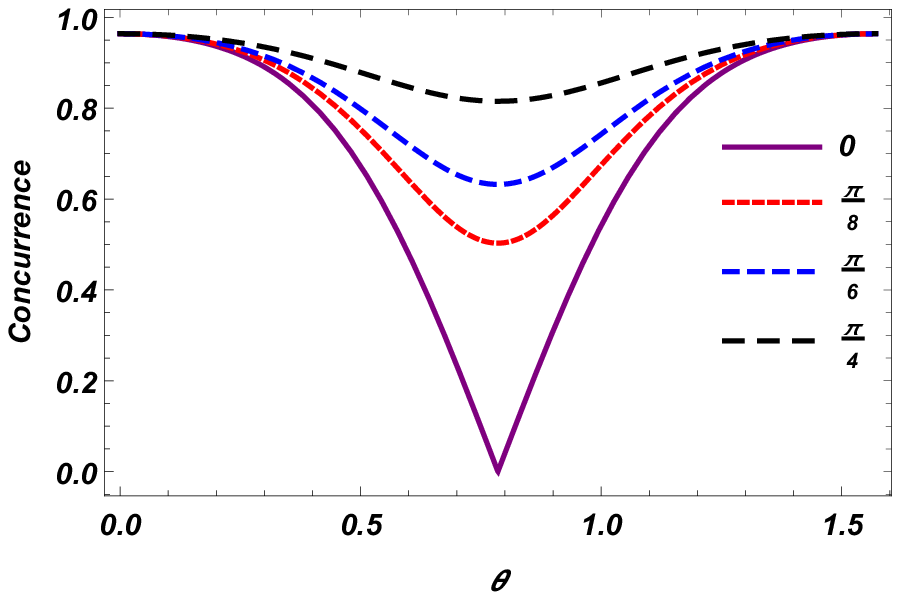}
  \caption{ }
  \label{both1}
\end{subfigure}
\begin{subfigure}{.5\textwidth}
  \includegraphics[width=1\linewidth]{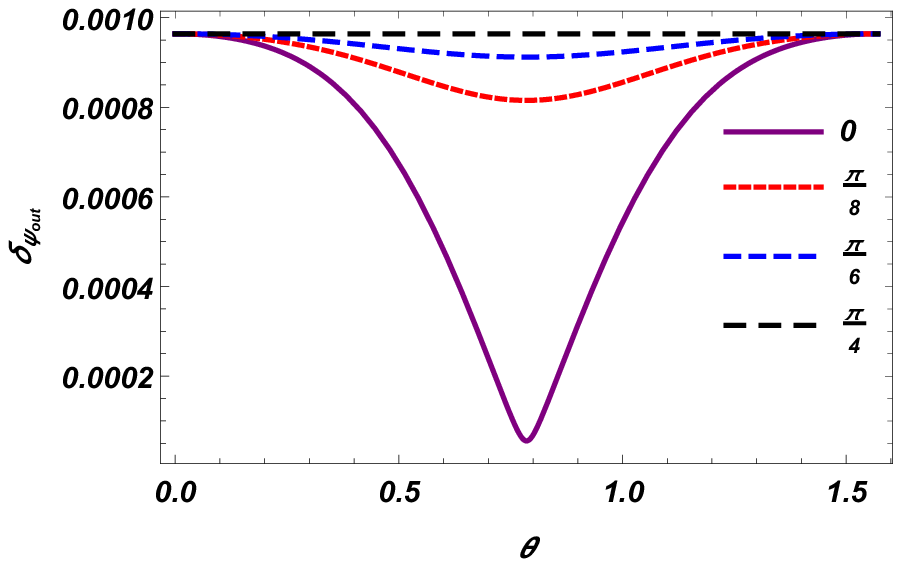}
  \caption{}
    \label{both2}
\end{subfigure}
\caption{The concurrence (a) and the NWF (b) versus rotator's angle $\theta$ for $\phi_1\in \lbrace 0,\dfrac{\pi}{8},\dfrac{\pi}{6},\dfrac{\pi}{4}\rbrace$ and $\phi_2=0$.}
\label{outfig}
\end{figure}

Now let as assume that, the state \eqref{s1} pass by $\textbf{C}\left(\phi_2\right)\textbf{R}\left(\theta\right)\textbf{C}\left(\phi_1\right)$ device (Compensator-Rotationer-Compensator device), where $\textbf{C}\left(\phi\right)=\textnormal{Exp}\left(i\dfrac{\phi}{2}\hat{S}_1\right)$
is the application of phase shift $\phi$ between the horizontal and vertical modes and $\textbf{R}\left(\theta\right)=\textnormal{Exp}\left(i\dfrac{\theta}{2}\hat{S}_3\right)$ is a geometric rotation by angle $\theta$ in the polarization. The quantum state at the output is given by 
\begin{footnotesize}
\begin{align}\label{outstate}
\ket*{\psi_{out}}=N_1&\biggl(\ket*{\beta\sin\left(\theta\right)e^{i\left(\phi_2-\phi_1\right)\slash 2}+\alpha\cos\left(\theta\right)e^{i\left(\phi_2+\phi_1\right)\slash 2}}\ket*{\beta\cos\left(\theta\right)e^{-i\left(\phi_2+\phi_1\right)\slash 2}-\alpha\sin\left(\theta\right)e^{-i\left(\phi_2-\phi_1\right)\slash 2}}\nonumber\\
& +\ket*{\alpha\sin\left(\theta\right)e^{i\left(\phi_2-\phi_1\right)\slash 2}+\beta\cos\left(\theta\right)e^{i\left(\phi_2+\phi_1\right)\slash 2}}\ket*{\alpha\cos\left(\theta\right)e^{-i\left(\phi_2+\phi_1\right)\slash 2}-\beta\sin\left(\theta\right)e^{-i\left(\phi_2-\phi_1\right)\slash 2}}
 \biggr).
\end{align}
\end{footnotesize}The NWF and the concurrence of the output state \eqref{outstate} are plotted in the Figure \eqref{outfig} versus $\lvert\alpha\rvert^2$ for $\lvert\beta\lvert^2=2$.

\section{Discussion and conclusion}\label{discussion}

In this section, we will discuss comparatively the behavior of the NWF as a measure of entanglement and the concurrence under a polarization converter device, $\textbf{C}\left(\phi_2\right)\textbf{R}\left(\theta\right)\textbf{C}\left(\phi_1\right)$. For this purpose we chose to plot the NWF and the concurrence for the input and output states $\ket*{\psi_1}$ and $\ket*{\psi_{out}}$ (\eqref{s1} and \eqref{outstate}).\\
Figure \eqref{outfig} shows the entanglement and the NWF in the input state \eqref{s1} that are dependent to the parameters $\theta$ and $\phi$ according to $\lvert\alpha\rvert$ and $\lvert\beta\rvert$. We see that the entanglement increases with increasing values of the parameter $\alpha$ to reach its maximum when $\lvert\alpha\rvert^2\geq1.5$  that is a result of the fact that at the limit of large values of the parameter $\lvert\alpha-\beta\rvert$ the coherent states $\ket*{\alpha}$ and $\ket*{\beta}$ become orthogonal. Thus the behavior of the bi-mode superposed coherent state is, as expected, exactly that of the Bell state.\\
After the state passed through the \textbf{CRC} device, we show in Figure \eqref{outfig} its entanglement as a function of $\theta$   for different values of  $\phi_1$ fixing $\lvert\alpha-\beta\rvert^2=4$. It is interesting to say that the rotation of $\dfrac{\pi}{4}$ applied on the input state $\ket*{\psi_1}$ destroys completely the entanglement. This implies that, the $\textbf{R}\left(\theta\right)$ device can be a perfect entangler$\slash$disentangler gate.\\
In figure \eqref{outfig}, the NWF is plotted as a function of $\theta$ for  $\lvert\alpha-\beta\rvert^2=4$ for different values of $\phi_1$. For a specific  values of $\phi_1$, we see that, the NWF decrease with increasing values of $\theta$ to reach its minimum and vanish for   $\phi_1=\dfrac{\pi}{4}$. Then, it increases again with increasing values of $\theta$. This allows to show that the NWF and the concurrence behave identically and they have  the same inflection points which does confirm that the NWF is a true measure of entanglement in non-Gaussian states.\\

As conclusion, in this paper we have studied the behavior of the entanglement and the polarization degree in superposition of two-mode coherent states. We have confirm that the NWF can be used as a good quantifier of entanglement in non-Gaussian systems.\\
As matter of fact, it turn out that the volume of the negative part of Wigner function is in fact a best quantifier of bipartite entanglement in non-Gaussian systems.\\
This work allows as to describe the Wigner function and the polarization of superposition of two-mode coherent states and the important use of the WF to study the entanglement in non-Gaussian systems. Consequently, the  NWF can be considered as a measure of entanglement in non-Gaussian systems. We believe that this result will be efficienct in quantum information theory, mostly in quantum computing \cite{forcer2002superposition}, because the Wigner function can be measured experimentally, \cite{smithey1993measurement,banaszek1999direct}, including the measurements of its negative values \cite{kenfack2004negativity}. The interest point on such experiments has triggered a search for operational definitions of the Wigner functions, based on experimental setup \cite{lougovski2003fresnel,leonhardt1997measuring}. It does represent a major step forward in the detection and the quantification of non-Gaussian entanglement in bipartite systems.

\bibliographystyle{ieeetr} 
\bibliography{bibfile}

\begin{appendices}
\numberwithin{equation}{section}
\section{Average values and variances of quantum Stokes parameters and Q-function:}
\label{average}
For each Quantum Stokes parameter $\hat{S}_i$ the variance is defined by $V_i=\left\langle\hat{S}_i^2\right\rangle-\left\langle\hat{S}_i\right\rangle^2$, where the averages of the quantum Stokes parameters and of their squared values in the state $\ket*{\psi_\pm}$ defined in \eqref{super} are:
\begin{equation}
\left\langle\hat{S}_1\right\rangle=\lvert N\rvert^2 \Big(\left(\lvert\alpha\rvert^2 -\lvert\beta\rvert^2\right)+\left(\lvert\gamma\rvert^2 -\lvert\lambda\rvert^2\right)+\big[\left(\alpha^*\gamma-\beta*\lambda\right)+ \left(\alpha\gamma^*-\beta\lambda^*\right)\big]\delta\Big)
\end{equation}
\begin{equation}
\left\langle\hat{S}_2\right\rangle=\lvert N\rvert^2 \Big(\left(\alpha^*\beta+\alpha\beta^*\right)+\left(\gamma^*\lambda+\gamma\lambda^*\right)+\big[\left(\alpha^*\lambda+\gamma\beta^*\right)+\left(\gamma^*\beta+\alpha\gamma^*\right)\big]\delta\Big)
\end{equation}
\begin{equation}
\left\langle\hat{S}_3\right\rangle=\lvert N\rvert^2 \Big(\left(\alpha\beta^*-\alpha^*\beta\right)+\left(\gamma\lambda^*-\gamma
^*\lambda\right)+\big[\left(\gamma\beta^*-\alpha^*\lambda\right)+\left(\alpha\lambda^*-\beta\gamma^*\right)\big]\delta\Big)
\end{equation}

\begin{equation}
\left\langle\hat{S}_1^2\right\rangle=\lvert N\rvert^2\left\lbrace \begin{array}{cc}
\lvert\alpha\rvert^2+\lvert\beta\rvert^2+\lvert\gamma\rvert^2+\lvert\lambda\rvert^2+\left(\lvert\alpha\rvert^2 -\lvert\beta\rvert^2\right)^2+\left(\lvert\gamma\rvert^2 -\lvert\lambda\rvert^2\right)^2+\\
\big[\alpha^*\gamma+\beta^*\lambda+\alpha\gamma^*+\beta\lambda^*+\left(\alpha^*\gamma-\beta^*\lambda\right)^2+\left(\alpha\gamma^*-\beta\lambda^*\right)^2\big]\delta
\end{array} \right\rbrace
\end{equation}

\begin{equation}
\left\langle\hat{S}_2^2\right\rangle=\lvert N\rvert^2\left\lbrace \begin{array}{cc}
\lvert\alpha\rvert^2+\lvert\beta\rvert^2+\lvert\gamma\rvert^2+\lvert\lambda\rvert^2+2\left(\lvert\alpha\rvert^2 \lvert\beta\rvert^2+\lvert\gamma\rvert^2\lvert\lambda\rvert^2\right)+\left(\alpha^*\beta\right)^2+\left(\alpha\beta^*\right)^2\left(\gamma^*\lambda\right)^2+\\
\left(\gamma\lambda^*\right)^2+\big[\alpha^*\gamma+\beta^*\lambda+\alpha\gamma^*+\beta\lambda^*+\left(\alpha^*\lambda+\beta^*\gamma\right)^2+\left(\gamma^*\beta+\alpha\lambda^*\right)^2\big]\delta

\end{array} \right\rbrace
\end{equation}

\begin{equation}
\left\langle\hat{S}_3^2\right\rangle=\lvert N\rvert^2\left\lbrace \begin{array}{cc}
\lvert\alpha\rvert^2+\lvert\beta\rvert^2+\lvert\gamma\rvert^2+\lvert\lambda\rvert^2+2\left(\lvert\alpha\rvert^2 \lvert\beta\rvert^2-\lvert\gamma\rvert^2\lvert\lambda\rvert^2\right)-\left(\alpha^*\beta\right)^2-\left(\alpha\beta^*\right)^2\left(\gamma^*\lambda\right)^2-\\
\left(\gamma\lambda^*\right)^2+\big[\alpha^*\gamma+\beta^*\lambda+\alpha\gamma^*+\beta\lambda^*-\left(\alpha^*\lambda-\beta^*\gamma\right)^2-\left(\gamma^*\beta-\alpha\lambda^*\right)^2\big]\delta
\end{array} \right\rbrace
\end{equation}
where $\delta=\textnormal{exp}\big[\alpha^*\gamma+\beta^*\lambda-\left(\lvert\alpha\rvert^2+\lvert\beta\rvert^2+\lvert\gamma\rvert^2+\lvert\lambda\rvert^2\right)\slash2\big]$ and the Q-function of the same state (\eqref{super}) is 
\begin{equation}
Q\left(\theta,\phi\right)=\dfrac{\lvert N\rvert^2}{4\pi}\left\lbrace
\begin{array}{cc}
e^{-\left(\lvert\alpha\rvert^2+\lvert\beta\rvert^2\right)}\Big(1+z_1\Big)e^{z_1}+e^{-\left(\lvert\gamma\rvert^2+\lvert\lambda\rvert^2\right)}\Big(1+z_2\Big)e^{z_2}+\\
e^{-\left(\lvert\alpha\rvert^2+\lvert\beta\rvert^2+\lvert\gamma\rvert^2+\lvert\lambda\rvert^2\right)\slash 2}\left[\Big(1+z_{12}\Big)e^{z_{12}}+\Big(1+z_{12}^*\Big)e^{z_{12}^*}\right]
\end{array}\right\rbrace
\end{equation}

\end{appendices}

\end{document}